# Detection of a Sparse Change in High-Dimensional Time Series

Jingyan Huang      jingyan_huang@u.nus.edu

July 29, 2025

**Abstract**

Consider the detection of a sparse change in high-dimensional time-series. We introduce Sparsity Likelihood-based (SL-based) score and the change-points detection procedure in multivariate normal model with general covariance structure. SL-based algorithm is proved to achieve that supremum of error probabilities converges to 0. We run the simulation studies for SL-based algorithm and also illustrate its applications to a S&P500 dataset.

## 1  Introduction

We consider $N$ time series and each has a length of $T$. A problem considered is to detect the number and locations of mean changes when the changes are sparse across the time series. The settings of temporal dependence have been considered in (6), (7), and (9).

Change-point detection consists of $N$ independent change-point detection problems. We solve these $N$ problems together by summing $N$ appropriate transformations of p-values to help boosting detection power of change-point detection algorithms. The score $Z_t^n$ at location $t$ on the nth sequence is defined as the normalization of mean difference of the left and right intervals at location $t$ on the nth sequence. A common strategy to handle sparsity is to sum up scores across sequences after subjecting the scores to thresholding or penalizations. The Sparsified Binary Segmentation (SBS) (Cho and Fryzlewicz (6)), the double CUSUM (DC) (Cho (7)), the Informative Sparse Projection (INSPECT) (Wang and Samworth (9)) and the scan algorithm of Enikeeva and Harachaoui (8) are the algorithms that employ this strategy. This strategy helps to surpress noise from the sequences without change points.

The higher-criticism (HC) test statistic (Tukey (3)) that uses multiple thresholds for sparse mixture detection, is proposed to check for significantly large number of small p-values. The p-values are calculated from the normalized scores for mean difference of left and right intervals at location $t$ on the nth sequence. The HC test statistic was shown by Donoho and Jin (2) to be optimal in the detection of a sparse normal mixture. The Berk-Jones test statistic by (1) was shown to achieve optimality in sparse normal mixture detection. A sparse likelihood score by (10) is also shown to be optimal in the detection of a sparse normal mixture.

A sparsity likelihood procedure by (10) is introduced to detect the mean changes sparse across the sequence with both spatial independence and temporal independence. The proposed method can handle data types varying across sequences since the sparsity likelihood scores are transformations of p-values. Furthermore, the sparsity likelihood procedure adopts a systematic two-step approach. The procedure is computationally efficient.

In this work, we consider an extended sparsity likelihood procedure, called SL-based, to detect mean changes sparse across multiple series of gaussian autoregressive process of order one. After normalizing the score for mean difference and combining p-values using sparsity likelihood score, SL-based employs the systematic two-step approach as in (10) to make the procedure computationally efficient. *The innovations of SL-based algorithm are that $Var(\bar{X}_{tu}^n - \bar{X}_{st}^n)$ in normalization for temporal dependence structure are different from that in temporal independence structure.* Furthermore, in this work, the asymptotic detectable boundaries at different levels of sparsity are provided, which are not shown in other literature on change-point detection in high-dimensional time-series. The SL-based procedure is also shown to achieve successful detection at these detectable boundaries.

The outline of this paper is as follows. In Section 2 we introduce SL-based score and the change-points detection procedure in multivariate normal model with general covariance structure. *In Section 3 we show that the SL-based algorithm achieves that supremum of error probabilities converges to 0.* In Section 4 we perform numerical studies on the SL-based algorithm. *In Section 5 we discuss potential extensions or limitations of the algorithm. In Appendix we prove Theorem 1.*

## 2 Multivariate normal model with general covariance structure and change-point detection

Consider $N$ sequences of length $T$. Let $X_t^n$ denote the $t$th observation of the $n$th sequence. Let $\mathbf{X}_1, \cdots, \mathbf{X}_T$ be $N-$dimensional random vectors sampled from

$$\mathbf{X}_t \sim MVN(\boldsymbol{\mu}_t, I_N), 1 \leq t \leq T, \tag{1}$$

where $\boldsymbol{\mu}_t = [\mu_t^1, \cdots, \mu_t^N]'$ is a deterministic mean vector.

We allow for temporal dependence but spatial independence. That means, each sequence generates independently but the random vectors $\mathbf{X}_1, \cdots, \mathbf{X}_T$ are not necessarily independent. We also assume the covariance structure of the random vectors $\mathbf{X}_1, \cdots, \mathbf{X}_T$ is known and does not depend on $n$. To be specific, $Cov([X_{s+1}^n, \cdots, X_t^n]') = \boldsymbol{\Sigma}_{s,t}$, which is a $(t-s) \times (t-s)$ matrix known for all $n$ and $s, t$ such that $0 \leq s < t \leq T$.

We consider the detection and estimation of the change-point set

$$\boldsymbol{\tau} := \{t : \mu_t^n \neq \mu_{t+1}^n \text{ for some } n\}. \tag{2}$$

To check for a mean change on the $n$th sequence at location $t$, we can select $s < t < u$ and consider the p-value

$$\begin{aligned} p_{stu}^n &= 2\Phi(-|Z_{stu}^n|), \\ \text{with } Z_{stu}^n &= \frac{\bar{X}_{tu}^n - \bar{X}_{st}^n}{\sqrt{a}}, \end{aligned} \tag{3}$$

where $a = Var(\bar{X}_{tu}^n - \bar{X}_{st}^n) = \frac{2\sum_{t<i<j\leq u}\sigma_{ij}+u-t}{(u-t)^2} + \frac{2\sum_{s<i<j\leq t}\sigma_{ij}+t-s}{(t-s)^2} - \frac{2\sum_{s<i\leq t<j\leq u}\sigma_{ij}}{(u-t)(t-s)}$, and $\sigma_{ij} = Cov(X_i^n, X_j^n)$, for all $n$.

Let $f_1(p) = \frac{1}{p(2-\log p)^2} - \frac{1}{2}$ and $f_2(p) = \frac{1}{\sqrt{p}} - 2$. Note that $\int_0^1 f_i(p)dp = 0$ for $i = 1$ and $2$. Let $\lambda_1 \geq 0$ and $\lambda_2 > 0$.

Define the sparsity likelihood score

$$\begin{aligned} \ell_N(\mathbf{p}) &= \sum_{n=1}^N \ell(p^n), \\ \text{where } \ell(p) &= \log\left(1 + \frac{\lambda_1 \log N}{N} f_1(p) + \frac{\lambda_2}{\sqrt{N \log N}} f_2(p)\right). \end{aligned} \tag{4}$$

In the sparsity likelihood algorithm we combine these p-values using $\ell_N(\mathbf{p}_{stu})$, where $\mathbf{p}_{stu} = (p_{stu}^1, \ldots, p_{stu}^N)$. For this algorithm, extensions to some other dis-tributions are possible but technically involved and we do not pursue them here.

Let the penalized sparsity likelihood scores be defined as

$$\ell_N^{\text{pen}}(\mathbf{p}_{stu}) = \ell_N(\mathbf{p}_{stu}) - \log(\tfrac{T}{4}(\tfrac{1}{t-s} + \tfrac{1}{u-t})). \tag{5}$$

To implement SL-based algorithm, we firstly zone out the $(s, t, u)$ that are evaluated by following the approximating set idea of (17), and to precisely locate the change-points using the CUSUM-like procedure of Wild Binary Segmentation (WBS) only when a change-point has been indicated during the first step.

Let $1 \leq h_1 < h_2 < \cdots$ and $1 \leq d_1 < d_2 < \cdots$ be integer-valued sequences with $h_i \geq d_i$ for all $i$. Let $K_i(g) = \lfloor \frac{g-1}{d_i} \rfloor$, where $\lfloor \cdot \rfloor$ denotes the greatest integer function.

Define

$$\begin{aligned}
\mathcal{A}_i(g) &= \{(s(ik), t(ik), u(ik)) : 1 \leq k \leq K_i(g)\}, \\
\text{with } s(ik) &= \max(0, kd_i - h_i), \; t(ik) = kd_i, \text{ and } u(ik) = \min(kd_i + h_i, g).
\end{aligned}$$

The elements of $\mathcal{A}_i(g)$ are the indices where sparsity scores for windows of length $h_i$ are computed. For $g \geq 1$, let $i_g = \max\{i : h_i + d_i \leq g\}$. The detection of change-points within $\mathbf{X}_{b:e}$, with window lengths at least $h_{i_0}$, is as follows.

**function** SL-based-estimate(c,$i_0$,b,e)
    $\mathbf{X} \leftarrow \mathbf{X}_{b:e}$
    $g \leftarrow e - b + 1$
    **for** $i = i_0, \ldots, i_g$
        **if** $\max_{1 \leq k \leq K_i(g)} \ell_N^{\text{pen}}(\mathbf{p}_{s(ik),t(ik),u(ik)}) \geq c$ **then**
            $j \leftarrow \text{argmax}_{k: 1 \leq k \leq K_i(g)} \ell_N^{\text{pen}}(\mathbf{p}_{s(ik),t(ik),u(ik)})$
            $\widehat{\tau} \leftarrow [\text{argmax}_{t: s(ij) < t < u(ij)} \ell_N^{\text{pen}}(\mathbf{p}_{s(ij),t,u(ij)})] + b - 1$

**output** $(\widehat{\tau}, i)$

                    **stop**

                **end if**

        **end for**

    **output** $(0,0)$

**end function**

*In SL-based-estimate, $i_0$ refers to the index of the smallest window length and $c$ refers to the threshold of testing penalized sparsity likelihood scores. $b$ and $e$ refer to data range of the detection of change-points.* The recursive segmentation algorithm for the computation of the estimated change-point set $\widehat{\boldsymbol{\tau}}$ is as follows. We initialize with parameters $(c, 1, 1, T, \emptyset)$, where $\emptyset$ denotes the empty set.

**function** SL-based-detect$(c, i_0, b, e, \widehat{\boldsymbol{\tau}})$

    $(\widehat{\tau}, i) \leftarrow$ SL-based-estimate$(c, i_0, b, e)$

    **if** $\widehat{\tau} > 0$ **then**

        $\widehat{\boldsymbol{\tau}} \leftarrow \widehat{\boldsymbol{\tau}} \cup \{\widehat{\tau}\}$

        $\widehat{\boldsymbol{\tau}} \leftarrow$ SL-based-detect$(c, i, b, \widehat{\tau}, \widehat{\boldsymbol{\tau}})$

        $\widehat{\boldsymbol{\tau}} \leftarrow$ SL-based-detect$(c, i, \widehat{\tau}, e, \widehat{\boldsymbol{\tau}})$

    **end if**

    **output** $\widehat{\boldsymbol{\tau}}$

**end function**

REMARKS 1. The change point detection procedure on sparsity likelihood score performs better compared to other competitors.

The proposed procedure can be extended to some other types of distribution changes and it can handle data types varying across sequences since the likelihood scores are transformations of p-values.

The proposed algorithm here has two steps in the identification of change-points. The first detection screening step applies the Screening and Ranking (SaRa) idea of Niu and Zhang (13) which evaluates a large amount of segments. The second estimation step applies the CUSUM-like procedure of Wild Binary Segmentation (WBS) of Fryzlewicz (14).

The two-step procedure is computationally efficient since the computationally intensive estimation step is only applied when a change-point has been detected during the first step. In contrast for WBS the estimation step is applied on a large number of randomly generated segments.

In addition to computational savings, the two-step procedure achieves that *supremum of error probabilities converges to 0 not only at all levels of sparsity, but also at all orders of change magnitudes by incorporating multi-scale penalization terms* similar to those used in Dümbgen and Spokoiny (15) and the SMUCE algorithm of Frick, Munk and Sieling (16).

## 3 Change-point detection under an AR(1) process

*Consider multivariate normal model with covariance structure in* (1). *Write* $K(u) = cov(\mathbf{X}_t, \mathbf{X}_{t+u})$, *where $K(u)$ is a diagonal matrix for every $u$.*

We define a constant $B(h)$ that depends on $h$ and the constant is related to the $Var(\bar{X}_{tu}^n - \bar{X}_{st}^n)$, where $t - s = u - t = h$. $B(h)$ is defined as

$$B(h) = ||2hK(0) + 4(h-i+1)\sum_{u=1}^{h-1} K(u) - 2hK(h) - 2\sum_{u=1}^{h-1}\Big(K(u) + K(2h-u)\Big)||_{op}, \quad (6)$$

for all $h \leq \frac{T}{2}$. And $||A||_{op} = \max_i \sigma_i(A)$ is denoted as the operator norm of matrix $A$, where $\sigma_1(A), \cdots, \sigma_{min(M,N)}(A)$ are the singular values of matrix

$A = (A_{ij}) \in R^{M \times N}$.

We consider $T \to \infty$ as $N \to \infty$ such that

$$\log T \sim N^\zeta \text{ for some } 0 < \zeta < 1. \tag{7}$$

In this section, we derive the asymptotic detectable bounds for different levels of sparsity and show the SL-based algorithm could achieve successful detection at these detectable boundaries in Theorem 1. We consider here the successful detection of change-points.

Recall that $i_T = \max\{i : h_i + d_i \leq T\}$. We apply the SL algorithm with $d_i$ and $h_i$ satisfying

$$\frac{h_{i+1}}{h_i} \to 1 \text{ and } d_i = o(h_i) \text{ as } i \to \infty, \tag{8}$$

$$\log \Big( \sum_{i=1}^{i_T} \frac{h_i}{d_i} \Big) = o(\log T) \text{ as } T \to \infty, \tag{9}$$

and critical values $c_T$ satisfying

$$c_T = o(\log T) \text{ and } c_T - \log \Big( \sum_{i=1}^{i_T} \frac{h_i}{d_i} \Big) \to \infty \text{ as } T \to \infty. \tag{10}$$

For the SL-based algorithm we select parameters

$$\lambda_1 > 0 \text{ and } \lambda_2 = \sqrt{\frac{\log T}{\log \log T}}. \tag{11}$$

(8) is satisfied when

$$h_i \sim \exp(\tfrac{i}{\log i}) \text{ and } d_i \sim \tfrac{h_i}{i} \text{ as } i \to \infty.$$

Moreover (9) is satisfied as

$$\log \Big( \sum_{i=1}^{i_T} \tfrac{h_i}{d_i} \Big) \sim 2 \log i_T \sim 2 \log \log T.$$

The choice of $\lambda_i$ in (11) and the role of $f_i$ in the detection of change-points are explained as follows. By the approximation $\log(1+x) \sim x$ as $x \to 0$ and $\log(1+x) \sim \log x$ as $x \to \infty$,

$$\ell_N(p) \sim \begin{cases} \log(\tfrac{1}{Np}) \text{ for } p = o(\tfrac{1}{N \log N}), \\ \tfrac{\lambda_2}{\sqrt{pN \log N}} \text{ for } p = o(1) \text{ but large compared to } \tfrac{\lambda_2^2}{N \log N}. \end{cases}$$

To show Theorem 1 i where the number of sequences undergoing change small compared to $\log T$, the first approximation is used. To show Theorem 1 ii where the number of sequences undergoing change large compared to $\log T$, the second approximation is used.

Instead of $\lambda_2 = \sqrt{\tfrac{\log T}{\log \log T}}$, SL-based algorithm can achieve the detectable boundaries more generally by letting

$$\lambda_2 = o(\sqrt{\log T}) \text{ with } \log \lambda_2 \sim \tfrac{1}{2} \log \log T.$$

Let

$$m_{j\Delta} = \#\{n : |\mu_{\tau_j+1}^n - \mu_{\tau_j}^n| \geq \Delta\}$$

be the number of sequences undergoing mean change of at least $\Delta$ at the $j$th change-point. Let

$$\begin{aligned} \Omega_0 &= \{\boldsymbol{\mu} : J = 0\}, \\ \Omega_1(\Delta, V, h) &= \{\boldsymbol{\mu} : \text{ there exists } j \text{ such that } \min(\tau_j - \tau_{j-1}, \tau_{j+1} - \tau_j) \geq h, \\ &\quad \text{ and } m_{j\Delta} \geq V\}, \end{aligned}$$

with the convention $\tau_0 = 0$ and $\tau_{J+1} = T$. We consider here the test of $H_0$: $\boldsymbol{\mu} \in \Omega_0$ versus $H_1$: $\boldsymbol{\mu} \in \Omega_1(\Delta, h, V)$. Consider the constant $\rho_Z(\beta, \zeta)$ defined as follows

$$\rho_Z(\beta, \zeta) = \begin{cases} \beta - \frac{1-\zeta}{2}, & \text{if } \frac{1-\zeta}{2} < \beta \leq \frac{3(1-\zeta)}{4}, \\ (\sqrt{1-\zeta} - \sqrt{1-\zeta-\beta})^2, & \text{if } \frac{3(1-\zeta)}{4} < \beta \leq 1-\zeta. \end{cases}$$

The tests are for the difference of means against a known baseline and there can be at most one change of mean in (18) and (19). There are no baselines and there can be multiple changes of mean along the sequences in the setting considered here.

**Theorem 1.** *Assume* (1), (6) *and* (7). *Let* $0 < \epsilon < 1$ *and* $\Delta > 0$ *fixed. The SL-based algorithm, with parameters satisfying* (8)–(11), *achieves*

$$\sup_{\boldsymbol{\mu} \in \Omega_0} P_{\boldsymbol{\mu}}(\text{Type I error}) + \sup_{\boldsymbol{\mu} \in \Omega_1(\Delta, V, h)} P_{\boldsymbol{\mu}}(\text{Type II error}) \to 0 \qquad (12)$$

*under any of the following conditions.*

i. *When* $V = o(\frac{\log T}{\log N})$ *and* $\frac{h}{B(h)} = 4(1+\epsilon)(\frac{\log T}{V\Delta^2})$.

ii. *When* $V \sim N^{1-\beta}$ *for some* $\frac{1-\zeta}{2} < \beta < 1-\zeta$ *and* $\frac{h}{B(h)} = 4(1+\epsilon)\rho_Z(\beta, \zeta)(\frac{\log N}{\Delta^2})$.

## 4 Numerical Study

In this section, we conduct the simulation studies of SL-based algorithm in Section 4.1 and the application of SL-based algorithms on $S\&P500$ dataset in Section 4.2.

## 4.1 Simulation

We extend the simulation set-up in Sections 5.1 and 5.3 of (9).

In the first study there is exactly one change-point $\tau_1 = 0.4T$. We consider the set-up $N, T \in \{500, 2000\}$, the number of sequences undergoing change-points $V = \{3, 22, 50, 500\}$ when $N = 500$ and $V = \{3, 45, 200, 2000\}$ when $N = 2000$.

Consider $\mu_t^n = 0$ for $t \leq \tau_1$ and all $n$. For $t > \tau_1$, let

$$\mu_t^n = \begin{cases} 1.2 \Big/ \sqrt{n \sum_{m=1}^{V} m^{-1}} & \text{if } n \leq V, \\ 0 & \text{if } n > V. \end{cases} \tag{13}$$

Consider the Gaussian $AR(1)$ process:

$$X_t^n = c + \varphi X_{t-1}^n + \varepsilon_t^n, \text{ for } 1 \leq t \leq T, 1 \leq n \leq N, \tag{14}$$

where $\varepsilon_t^n$ is a white noise process with zero mean and constant variance $\sigma_\varepsilon^2$. Assume that $\varphi = 1$ and $\sigma_\varepsilon^2 = 1$. $\ell_N^{\text{pen}}$ is the penalized sparsity score with $\lambda_1 = 1$ and $\lambda_2 = \sqrt{\frac{\log T}{\log \log T}}$.

We simulate the probabilities that $|\hat{\tau}_1 - \tau_1| \leq k$ for $k = 3$ and $10$, and compare SL-based algorithm against the INSPECT algorithm of (9). The comparisons in Table 1 show that SL-based algorithm performs well.

In the second simulation study there are three change-points $\boldsymbol{\tau} = (500, 1000, 1500)$. We consider $T = 2000, N = 200, V = 40$. Consider the Gaussian AR(1) process in (14) with $\varphi = 1$ and $\sigma_\varepsilon^2 = 1$.

At each change-point exactly 40 sequences undergo mean changes. Six scenarios are considered, corresponding to

$$\mu_{\tau_j+1}^{k(j-1)+n} - \mu_{\tau_j}^{k(j-1)+n} = r \Big/ \sqrt{n \sum_{m=1}^{40} m^{-1}},$$

| $T$ | $N$ | $V$ | SL-based | | INSPECT | |
|---|---|---|---|---|---|---|
| | | | $k=3$ | $k=10$ | $k=3$ | $k=10$ |
| 500 | 500 | 3 | *0.664* | *0.895* | 0.609 | 0.860 |
| | | 22 | *0.498* | *0.760* | 0.453 | 0.729 |
| | | 50 | *0.436* | *0.691* | 0.373 | 0.653 |
| | | 500 | *0.310* | *0.552* | 0.273 | 0.504 |
| 500 | 2000 | 3 | *0.643* | *0.899* | 0.492 | 0.766 |
| | | 45 | *0.361* | *0.625* | 0.232 | 0.454 |
| | | 200 | *0.231* | *0.461* | 0.164 | 0.358 |
| | | 2000 | *0.146* | *0.323* | 0.107 | 0.279 |
| 2000 | 500 | 3 | *0.707* | *0.930* | 0.690 | 0.923 |
| | | 22 | 0.649 | *0.896* | *0.650* | 0.895 |
| | | 50 | *0.601* | 0.867 | 0.596 | *0.868* |
| | | 500 | *0.489* | 0.781 | 0.482 | *0.783* |
| 2000 | 2000 | 3 | *0.703* | *0.924* | 0.672 | 0.909 |
| | | 45 | *0.579* | *0.844* | 0.546 | 0.824 |
| | | 200 | *0.487* | *0.765* | 0.450 | 0.734 |
| | | 2000 | *0.397* | *0.675* | 0.354 | 0.607 |

Table 1: The fraction of simulation runs (out of 1000) for which $\widehat{\tau}_1$ is within distance $k$ from $\tau_1$ for $k=3$ and 10. The same datasets are used to compare SL-based algorithm and INSPECT algorithm with $\tau_1 = 200$ for $T = 500$ and $\tau_1 = 800$ for $T = 2000$.

|           | SL-based | INSPECT | DC    | SBS  |
|-----------|----------|---------|-------|------|
| Threshold | 5.50     | 18.70   | 33.50 | 5.30 |

Table 2: Detection thresholds of SL-based, INSPECT, DC and SBS for which $\alpha = 0.05$.

where $1 \leq j \leq 3, 1 \leq n \leq 40$, $r = 0.8, 1$ and $k = 0, 20, 40$. For $k = 0$, the mean changes are within the same 40 sequences at all three change-points, whereas for $k = 40$ the mean changes at all three change-points are on distinct sequences. For $k = 20$, there is partial overlap of the sequences having mean changes at adjacent change-points. The number of estimated change-points over 100 simulated datasets on each sequence is recorded, as well as the adjusted Rand index (ARI), see (20) and (21), to measure the quality of the change-point estimation.

In the application of the SL-based algorithm, we select $h_1 = 1$ and $h_{i+1} = \lceil 1.1 h_i \rceil$ for $i \geq 1$, and $d_i = \lfloor h_i/i \rfloor$, for a total of $i_T = 61$ window lengths. We select parameters $\lambda_1 = 1$, $\lambda_2 = \sqrt{\frac{\log T}{\log \log T}} \doteq 1.94$.

We compare SL-based algorithm with INSPECT algorithm in (9), SBS algorithm by (6) and DC procedure by (7). For all algorithms, we apply detection thresholds for which $\alpha = P_{\boldsymbol{\mu}}(\hat{J} > 0) = 0.05$ when there are no change-points in all sequences. The detection thresholds are shown in Table 2.

Table 3 shows that compared to INSPECT, DC and SBS, SL-based has the highest average ARI for each of the six scenarios.

| $r$ | $k$ | Algorithm | # change-points | | | | | ARI |
|---|---|---|---|---|---|---|---|---|
| | | | 2 | 3 | 4 | 5 | 6 | |
| 1 | 0 | SL-based | 0 | 96 | 4 | 0 | 0 | *0.803* |
| | | INSPECT | 3 | 94 | 3 | 0 | 0 | 0.798 |
| | | DC | 1 | 82 | 15 | 2 | 0 | 0.782 |
| | | SBS | 2 | 95 | 3 | 0 | 0 | 0.781 |
| 0.8 | 0 | SL-based | 2 | 92 | 6 | 0 | 0 | *0.789* |
| | | INSPECT | 47 | 52 | 1 | 0 | 0 | 0.695 |
| | | DC | 15 | 69 | 13 | 1 | 2 | 0.728 |
| | | SBS | 22 | 72 | 6 | 0 | 0 | 0.723 |
| 1 | 20 | SL-based | 0 | 96 | 4 | 0 | 0 | *0.802* |
| | | INSPECT | 3 | 94 | 3 | 0 | 0 | 0.798 |
| | | DC | 2 | 89 | 8 | 1 | 0 | 0.788 |
| | | SBS | 2 | 87 | 11 | 0 | 0 | 0.771 |
| 0.8 | 20 | SL-based | 4 | 92 | 4 | 0 | 0 | *0.786* |
| | | INSPECT | 48 | 51 | 1 | 0 | 0 | 0.694 |
| | | DC | 19 | 75 | 5 | 1 | 0 | 0.739 |
| | | SBS | 21 | 68 | 11 | 0 | 0 | 0.722 |
| 1 | 40 | SL-based | 0 | 97 | 3 | 0 | 0 | *0.802* |
| | | INSPECT | 3 | 94 | 3 | 0 | 0 | 0.798 |
| | | DC | 1 | 88 | 10 | 1 | 0 | 0.787 |
| | | SBS | 2 | 96 | 2 | 0 | 0 | 0.779 |
| 0.8 | 40 | SL-based | 6 | 91 | 3 | 0 | 0 | *0.781* |
| | | INSPECT | 48 | 51 | 1 | 0 | 0 | 0.694 |
| | | DC | 17 | 76 | 6 | 1 | 0 | 0.735 |
| | | SBS | 24 | 70 | 6 | 0 | 0 | 0.717 |

Table 3: Number of change-points estimated and the average ARI over 100 simulated datasets by SL-based-detect, INSPECT, DC and SBS.

## 4.2 S&P500 Dataset

We further study the performance of the SL-based algorithm by applying it to the multiple time series of log returns of daily closing prices of S&P 500 stock market indices (The link to the dataset is https://finance.yahoo.com/lookup). It is common that the change-points are shared by only a subset of S&P 500 *stock market indices*. It is suitable to apply SL-based algorithm to this financial time series dataset. We consider S&P 500 stock market index in the period from 3 January 2007 to 14 April 2022, overlapping with the period of the financial crisis in 2008 and the COVID-19 recession starting from 2020. We have chosen only those component stocks that remained in the index over the entire period. The component stocks that are deviated from normality are also removed by checking the skewness of each series. The resulting time series is with $N = 294$ and $T = 3849$. *We conduct Mardia's test to check multivariate normality of remaining variables. The results show that p-value is 0.24, which is larger than* 0.05. *Mardia's test is not rejected at the significance level of* 0.05, *which means that it aligns with the assumption of multivariate normality. We further assess the validity of the assumption that each sequence is generated independently and follows a Gaussian AR(1) process. For independence of each stock component, we do the log transformation and first differencing of each sequence. Next, we check the independence of each sequence by calculating the correlation matrix of data sequences. The results show all correlations are close to zero, which suggests independence. Furthermore, we assess the validity that each stock component follows a Gaussian AR(1) process. Firstly, each data sequence is transformed and we check the stationarity of each data sequence. Secondly, we calculate autocorrelation Function (ACF) of each data sequence and check order of AR process, which is AR(1) process. Thirdly, maximum likelihood estimation (MLE) is applied to estimate the coefficients in*

| $t$ | Date | Historical event |
|---|---|---|
| 447 | 2008/10/9 | Around the date when the head of the International |
| 449 | 2008/10/13 | Monetary Fund (IMF) warned that the world financial system was teetering on the "brink of systemic meltdown". |
| 460 | 2008/10/28 | The 2008 U.S. presidential election was a significant event at that |
| 462 | 2008/10/30 | time, with candidates Barack Obama and John McCain campaigning intensively. On October 28, 2008, the campaign was in full swing, and the election was just a week away. |
| 481 | 2008/11/26 | Around the date when Economist Dean Baker observed: Tens of millions of homeowners who had substantial equity in their homes two years ago have little or nothing today. Businesses are facing the worst downturn since the Great Depression in September of 2008 (the Bankruptcy of Lehman Brothers). |
| 3319 | 2020/03/09 | European stock markets closed mostly up |
| 3320 | 2020/03/10 | while Asia-Pacific stock markets mostly closed down, |
| 3321 | 2020/03/11 | while the Dow Jones Industrial Average, NASDAQ Composite, and the $S\&P$ 500 all rose by more than 9%. |

Table 4: Summary of change-points detected from the component processed of S&P500

AR(1) process. Therefore, the assumption that each sequence is generated independently and follows a Gaussian AR(1) process is verified. Then we further estimate $c, \sigma_\varepsilon^2$, and $\varphi$ by ordinary least square estimation. In the application of the SL-based algorithm, we select $h_1 = 1$ and $h_{i+1} = \lceil 1.1 h_i \rceil$ for $i \geq 1$, and $d_i = \lfloor h_i / i \rfloor$, for a total of $i_T = 68$ window lengths. We select parameters $\lambda_1 = 1$, $\lambda_2 = \sqrt{\frac{\log T}{\log \log T}} \doteq 1.98$ and select critical value $c_T$ so that SL-based returns eight change-points. We apply SL-based to the entire $N = 294$ time series, the SL-based algorithm returns the change-points summarized(with close change-points summarized as a change-point) in Table 4, which also lists some historical events that occurred close to the detected change-points. Overall, the results in Table 4 support the use of the SL-based algorithm in this case study. R codes are available in https://github.com/JingyanJessica/DetectionOfASparseChangeInHighDimensionalTimeSeries.

# 5 Discussion

*The paper considers the multivariate normal model with covariance structure in (1) and assumes that different sequences are independent. The limitations are that the assumptions of multivariate normality and spatial independence are stringent for real-world applications. Future research include exploring relaxing these conditions and exploring more algorithms.*

*The result in Theorem 1 is asymptotic, and it claims that the supremum of error probabilities converges to 0. Future research also include the extensions of the results to AR models with higher orders such as AR(2), AR(3), etc.*

# 6 Disclosure statement

The author reports there are no competing interests to declare.

# 7 Appendix

## 7.1 Proof of Theorem 1

Since $c_N \to \infty$, by Markov's inequality $P_0(\text{Type I error}) \leq e^{-c_N} \to 0$. The proof of $P_{\mu_N}(\text{Type II error}) \to 0$ applies Lemmas 1 and 2 below. Their proofs are in (10).

**Lemma 1.** *Let* $\mathbf{q} = (q^1, \ldots, q^N)$, *with* $q^n \overset{i.i.d.}{\sim}$ Uniform(0,1). *For fixed* $\lambda_1 \geq 0$ *and* $\delta > 0$,

$$\sup_{\delta \leq \lambda_2 \leq \sqrt{N}} P(\ell_N(\mathbf{q}) \leq -\lambda_2^2 \sqrt{\log N}) \to 0.$$

**Lemma 2.** *For* $\lambda_1 > 0$ *fixed,* $\delta \leq \lambda_2 \leq \sqrt{N}$ *for some* $\delta > 0$ *and* $\xi_N = o(N^{-\eta})$

for some $\eta > 0$ such that $\xi_N \geq \frac{\lambda_2^2}{2N}$, for large $N$,

$$\inf_{\substack{(p,q):p \leq \xi_N, \\ q \geq \lambda_2^2/N, p \leq q/2}} [\ell_N(p) - \ell_N(q)] \geq \frac{\lambda_2}{4\sqrt{N\xi_N \log N}}.$$

PROOF OF THEOREM $1i.ii.$.

For $(s,t,u) \in \mathcal{A}_i(T)$, the penalty of the SL scores is

$$\log(\tfrac{T}{4}(\tfrac{1}{t-s} + \tfrac{1}{u-t})) \geq \log(\tfrac{T}{2h_i}).$$

Moreover $\#\mathcal{A}_i(T) \leq \frac{T}{d_i}$. Hence by markov's inequality $P_0(\ell_N(\mathbf{p}_{stu}) \geq c) \leq e^{-c}$ and $c_T - \log(\sum_{i=1}^{i_T} \frac{h_i}{d_i}) \to \infty$, for $\boldsymbol{\mu} \in \Omega_0$,

$$\begin{aligned}
P_{\boldsymbol{\mu}}(\text{Type I error}) &\leq \sum_{i=1}^{i_T} \sum_{(s,t,u) \in \mathcal{A}_i(T)} P_{\boldsymbol{\mu}}(\ell_N(\mathbf{p}_{stu}) \geq c_T + \log(\tfrac{T}{2h_i})) \\
&\leq \sum_{i=1}^{i_T} \tfrac{T}{d_i} \exp(-c_T - \log(\tfrac{T}{2h_i})) \\
&= 2e^{-c_T} \sum_{i=1}^{i_T} \tfrac{h_i}{d_i} \to 0.
\end{aligned}$$

Consider $\boldsymbol{\mu} \in \Omega_1(\Delta, h, V)$ and let $\tau_j$ be the change-point satisfying the conditions in the definition of $\Omega_1(\Delta, h, V)$. Let $Q^n = 1$ if $|\mu_{\tau_j+1}^n - \mu_{\tau_j}^n| \geq \Delta$ and $Q^n = 0$ otherwise. We assume without loss of generality that $0 < \epsilon < 1$.

To aid in the checking of the proof of Theorem 1, we provide here the key ideas. Let $j$ be such that

$$\min(\tau_j - \tau_{j-1}, \tau_{j+1} - \tau_j) \geq h \text{ and } m_{j\Delta} \geq V.$$

Consider $\Delta > 0$ fixed and $V \sim N^{1-\beta}$ for some $\frac{1-\zeta}{2} < \beta < 1 - \zeta$. Since $h \to \infty$, it follows from (6) and (8) that for *sufficiently large* $N$ we are able to find $(s, t, u) = (s(ik), t(ik), u(ik))$ close to $(\tau_j - h, \tau_j, \tau_j + h)$ such that

$$E_{\boldsymbol{\mu}} Z_{stu}^n \geq [1 + o(1)]\Delta \sqrt{\tfrac{h}{2B(h)}} \text{ for } n \text{ satisfying } |\mu_{\tau_{j+1}}^n - \mu_{\tau_j}^n| \geq \Delta. \tag{15}$$

Recall that $p_{stu}^n = 2\Phi(-|Z_{stu}^n|)$ and let $q_{stu}^n = \Phi(-|Z_{stu}^n| + E_{\boldsymbol{\mu}} Z_{stu}^n) + \Phi(-|Z_{stu}^n| - E_{\boldsymbol{\mu}} Z_{stu}^n)$. Let

$$\Gamma = \{n : |Z_{stu}^n| \geq \sqrt{2\omega \log N},\ q_{stu}^n \geq N^{\zeta-1},\ |\mu_{\tau_{j+1}}^n - \mu_{\tau_j}^n| \geq \Delta\}, \tag{16}$$

with $\omega = 1 - \zeta$ when $\tfrac{3(1-\zeta)}{4} < \beta < 1 - \zeta$ and $\omega = 4(\beta - \tfrac{1-\zeta}{2})$ when $\tfrac{1-\zeta}{2} < \beta \leq \tfrac{3(1-\zeta)}{4}$. It follows from Lemmas 1 and 2 that with probability tending to 1,

$$\begin{aligned}
\ell_N(\mathbf{p}_{stu}) &\geq \ell_N(\mathbf{q}_{stu}) + (\#\Gamma)\tfrac{\lambda_2}{4\sqrt{N\xi_N \log N}} \\
&\geq -\lambda_2^2 \sqrt{\log N} + (\#\Gamma)\tfrac{\lambda_2}{4\sqrt{N\xi_N \log N}} \text{ for } \xi_N = N^{-\omega}.
\end{aligned}$$

Since the penalty $\log(\tfrac{T}{4}(\tfrac{1}{t-s} + \tfrac{1}{u-t})) \leq \log T \sim N^\zeta$, $c_T = o(\log T)$ and $\lambda_2 \sim \tfrac{N^{\frac{\zeta}{2}}}{\sqrt{\zeta \log N}}$, to show $P_{\boldsymbol{\mu}}(\ell_{stu}^{\text{pen}}(\mathbf{p}) \geq c_T) \to 1$, it suffices to show that there exists $\delta > 0$ such that

$$E_{\boldsymbol{\mu}}(\#\Gamma) \gtrsim \begin{cases} N^{\zeta+\delta} & \text{if } \tfrac{3(1-\zeta)}{4} < \beta < 1 - \zeta, \\ N^{\frac{3}{2} - 2\beta - \frac{\zeta}{2} + \delta} & \text{if } \tfrac{1-\zeta}{2} < \beta \leq \tfrac{3(1-\zeta)}{4}. \end{cases} \tag{17}$$

PROOF OF THEOREM 1$i$.

Consider $V = o(\tfrac{\log T}{\log N})$. Since $\tfrac{h}{B(h)} = 4(1+\epsilon)(\tfrac{\log T}{\Delta^2 V}) \to \infty$, $\tfrac{h_{i+1}}{h_i} \to 1$ and $d_i = o(h_i)$, for large $T$ there exists

$$\frac{h_i}{B(h_i)} \geq 4(1+\epsilon)^{\frac{1}{2}}(\tfrac{\log T}{\Delta^2 V})$$

such that for all $\boldsymbol{\mu} \in \Omega_1(h, \Delta, V)$, there exists $k$ satisfying

$$\tau_{j-1} < s(ik) < u(ik) < \tau_{j+1} \text{ and } |t(ik) - \tau_j| \leq \tfrac{d_i}{2}. \tag{18}$$

Hence when $Q^n = 1$ and by (6),

$$|E_{\boldsymbol{\mu}} Z^n_{stu}| \geq \Delta(1 - \tfrac{d_i}{2h_i})\sqrt{\tfrac{h_i}{2B(h_i)}} \geq \sqrt{2(1+\epsilon)^{\frac{1}{3}} V^{-1} \log T}, \tag{19}$$

where $(s, t, u) = (s(ik), t(ik), u(ik))$.

Let $\Gamma = \{n : Q^n = 1, |Z^n_{stu}| \geq \sqrt{2(1+\epsilon)^{\frac{1}{4}}(\tfrac{\log T}{V})}\}$. Let $p^n_{stu} = 2\Phi(-|Z^n_{stu}|)$ and $q^n_{stu} = \Phi(-|Z^n_{stu}| + E_{\boldsymbol{\mu}} Z^n_{stu}) + \Phi(-|Z^n_{stu}| - E_{\boldsymbol{\mu}} Z^n_{stu})$. Since $q^n \overset{\text{i.i.d.}}{\sim}$ Uniform(0,1) and $\ell_N(1) \geq -1$ for large $N$, by Lemmas 1 and 2, with probability tending to 1,

$$\begin{aligned}
\ell_N(\mathbf{p}_{stu}) &\geq \ell_N(\mathbf{q}_{stu}) + (\#\Gamma)\left[\ell_N\left(2\Phi\left(-\sqrt{2(1+\epsilon)^{\frac{1}{4}} \tfrac{\log T}{V}}\right)\right) - 1\right] \quad (20)\\
&\geq -\lambda_2^2 \sqrt{\log N} + V[(1+\epsilon)^{\frac{1}{5}} \tfrac{\log T}{V} - \log N]\\
&\geq (1+\epsilon)^{\frac{1}{6}} \log T.
\end{aligned}$$

Since the penalty $\log(\tfrac{T}{4}(\tfrac{1}{t-s} + \tfrac{1}{u-t})) \leq \log T$ and $c_T = o(\log T)$, it follows that $P_{\boldsymbol{\mu}}(\ell_N^{\text{pen}}(\mathbf{p}_{stu}) \geq c_T) \to 1$.

PROOF OF THEOREM 1$ii$.. **Case 1**: $V \sim N^{1-\beta}$ for $\tfrac{3(1-\zeta)}{4} \leq \beta < 1 - \zeta$. Since $\tfrac{h}{B(h)}\Delta^2 = 4(1+\epsilon)(\sqrt{1-\zeta} - \sqrt{1-\zeta-\beta})^2 \log N$ and $d_i = o(h_i)$, for large $N$ there exists $i$ satisfying $h_i \geq (1+\epsilon)^{-\frac{1}{2}} h$ such that whenever $Q^n = 1$ and by (6),

$$\begin{aligned}
|E_{\boldsymbol{\mu}} Z^n| &\geq \Delta(1 - \tfrac{d_i}{2h_i})\sqrt{\tfrac{h_i}{2B(h_i)}} \geq \sqrt{2\nu \log N}, \quad (21)\\
\nu &= (1+\epsilon)^{\frac{1}{3}}(\sqrt{1-\zeta} - \sqrt{1-\zeta-\beta})^2,
\end{aligned}$$

with $(s, t, u) = (s(ik), t(ik), u(ik))$ for $k$ satisfying (18).

For $\Gamma$ defined in (16),

$$
\begin{aligned}
E_{\boldsymbol{\mu}}(\#\Gamma) &\geq V[\Phi\Big(-\sqrt{2(1-\zeta)\log N} + \sqrt{2\nu \log N}\Big) - N^{\zeta-1}] \\
&\gtrsim N^{1-\beta-(\sqrt{1-\zeta}-\sqrt{\nu})^2}(\log N)^{-\frac{1}{2}},
\end{aligned}
$$

and (17) follows from

$$
\sqrt{1-\zeta} > \sqrt{\nu} > \sqrt{1-\zeta} - \sqrt{1-\zeta-\beta}.
$$

**Case 2**: $V \sim N^{1-\beta}$ for $\frac{1-\zeta}{2} < \beta < \frac{3(1-\zeta)}{4}$. Since $\frac{h}{B(h)}\Delta^2 = 4(1+\epsilon)(\beta - \frac{1-\zeta}{2})\log N$, for large $N$ there exists $h_i \geq (1+\epsilon)^{-\frac{1}{2}}h$ such that whenever $Q^n = 1$ and by (6),

$$
\begin{aligned}
|E_{\boldsymbol{\mu}} Z^n_{stu}| &\geq \Delta(1 - \tfrac{d_i}{2h_i})\sqrt{\frac{h_i}{2B(h_i)}} \geq \sqrt{2\nu \log N}, \quad (22)\\
\nu &= (1+\epsilon)^{\frac{1}{3}}(\beta - \tfrac{1-\zeta}{2}),
\end{aligned}
$$

with $(s, t, u) = (s(ik), t(ik), u(ik))$ for $k$ satisfying (18).

For $\Gamma$ defined in (16),

$$
\begin{aligned}
E_{\boldsymbol{\mu}}(\#\Gamma) &\geq V[\Phi\Big(-2\sqrt{(2\beta-1+\zeta)\log N} + \sqrt{2\nu\log N}\Big) - N^{\zeta-1}] \\
&\gtrsim N^{1-\beta-(2\sqrt{\beta-\frac{1-\zeta}{2}}-\sqrt{\nu})^2}(\log N)^{-\frac{1}{2}},
\end{aligned}
$$

and (17) follows from

$$
2\sqrt{\beta - \tfrac{1-\zeta}{2}} > \sqrt{\nu} > \sqrt{\beta - \tfrac{1-\zeta}{2}}.
$$

# References

[1] Berk, R.H. and Jones, D.H. (1979), "Goodness-of-fit test statistics that dominate the Kolmogorov statistics." Z. Wahrsch. Verw. Gebiete, 47, 47-50.

[2] Donoho, D. and Jin, J. (2004), "Higher criticism for detecting sparse heterogeneous mixtures." The Annals of Statistics, 32, 962-994.

[3] Tukey, J.W. (1976), "T13 N: The higher criticism." Course Notes, Statistics 411. Princeton Univ.

[4] Ingster, Y.I. (1997), "Some problems of hypothesis testing leading to infinitely divisible distributions. " Mathematical Methods of Statistics, 6, 47-69.

[5] Ingster, Y.I. (1998), " Minimax detection of a signal for $\ell^n$ balls. " Mathematical Methods of Statistics, 7,401-428.

[6] Cho, H., and Fryzlewicz, P. (2015), "Multiple change-point detection for high dimensional time series via sparsified binary segmentation," Journal of the Royal Statistical Society: Series B (Statistical Methodology), 77, 475-507.

[7] Cho, H. (2016), "Change-point detection in panel data via double cusum statistic," Electronic Journal of Statistics, 10, 2000-2038.

[8] Enikeeva, F. and Harchaoui, Z. (2019), "High-dimensional change-point detection under sparse alternatives," The Annals of Statistics, 47, 2051-2079.

[9] Wang, T., and Samworth, R. (2018), "High dimensional change point estimation via sparse projection," Journal of The Royal Statistical Society, 80, 57-83.

[10] Hu, S.R., Huang, J.Y., Chen, H., and Chan, H.P. (2023), "Likelihood scores

for sparse signal and change-point detection," IEEE Transactions on Information Theory, 69, 4065-4080.

[11] Pilliat, E., Carpentier, A., and Verzelen, N. (2023), "Optimal multiple change-point detection for high-dimensional data," Electronic Journal of Statistics, 17, 1240-1315.

[12] Liu, H., Gao, C., and Samworth, R. (2021), "Minimax rate in sparse high-dimensional change-point detection," The Annals of Statistics, 49, 1081-1112.

[13] Niu, Y.S. and Zhang, H. (2012), "The screening and ranking algorithm to detect DNA copy number variation," The Annals of Applied Statistics, 6, 1306-1326.

[14] Fryzlewicz, P. (2014), "Wild binary segmentation for multiple change-point detection," The Annals of Statistics, 42, 2243-2281.

[15] Dümbgen, L. and Spokoiny, V.G. (2001), "Multiscale testing of qualitative hypotheses," The Annals of Statistics, 29, 124-152.

[16] Frick, K., Munk, A. and Sieling, H. (2014), "Multiscale change point inference," Journal of the Royal Statistical Society: Series B (Statistical Methodology), 76, 495-580.

[17] Walther, G. (2010), "Optimal and fast detection of spatial clusters with scan statistics," The Annals of Statistics, 38, 1010-1033.

[18] Chan, H.P., and Walther, G. (2015), "Optimal detection of multi-sample aligned sparse signals," The Annals of Statistics, 43, 1865-1895.

[19] Chan, H.P. (2017), "Optimal sequential detection in multi-stream data," The Annals of Statistics, 45, 2736-2763.

[20] Rand, W. M. (1971), "Objective criteria for the evaluation of clustering methods," Journal of the American Statistical Association, 66, 846-850.

[21] Hubert, L., and Arabie, P. (1985), "Comparing partitions," Journal of Classification, 2, 193-218.